\documentclass[12pt,notitlepage]{article}
\usepackage{amsfonts}
\usepackage{amssymb}
\usepackage{graphicx}
\usepackage{amsmath}
\usepackage{endnotes}
\usepackage{setspace}
\usepackage{setspace}
\usepackage{fancyvrb}
\usepackage{tikz}

\usepackage{hyperref}
\usepackage{amsmath}

\begin{document}

\title{Repeat Contact and the Spread of Disease: \\
    An Agent Model with Compartmental Solution} 
\author{Peter Cotton}
\date{\today}
\maketitle

\begin{abstract}
   Using a probability of novel encounter derived from a physical model, we augment the SIR compartmental model for disease spread. Scenarios with the same initial trajectories and identical $R_0$ values can diverge greatly depending on the speed at which our circles of acquaintances grow stale - leading to order of magnitude differences in final case counts. A momentum effect arises from variation in the mean time since infection, and this feeds back into new infection rate and faster decline in the late stages of an outbreak. Rapid extinction of an outbreak can occur in the early stages, but once this opportunity is missed the effect is diminished and then, only herd immunity can help. 
 \end{abstract}

\section{Motivation}

In an extensive survey of early stage epidemic growth rates, Chowell et al conclude ``more refined models are needed however, in particular to account for variation in the early growth dynamics of real epidemics'' \cite{Chowell2016MathematicalReview}. The authors suggest on phenomenological grounds that an adjustment to infection rates be made. Under the suggested change infection is proportional not only to the current populations of infected and susceptible people, but also to the following time dependent function:
\begin{equation}
\label{eqn:chowell}
    \beta(t) =  \beta_0 \left\{ (1-\phi)e^{-qt}+\phi \right\}
\end{equation}
The authors suggest that this attenuation of infection rate stands a much better chance of fitting empirical data from a wide variety of real world diseases, including those seeming to exhibit sub-exponential growth. 

As a general comment there is no a prior reason why growth should be exponential, whether growth refers to the growth of disease, the uptake of a particular rumor or for that matter, an increase in fish population \cite{Chen1992AData}. However it is perhaps instructive, as Chowell and co-authors suggest, to treat exponential growth (or equivalently proportional infection) as a starting point to be perturbed. 

In that spirit we consider the role played by contact repetition in the spread of infectious disease, an effect studied in simulations by Smieszek, Fiebig and Scholz \cite{Smieszek2009ModelsIncluded}. Ferrari, Perkins, Pomeroy et al advocate network models to capture this effect and explain different transmission rates \cite{Ferrari2011PathogensScaling}. It has elsewhere been argued that randomness of these networks plays an important role \cite{Miller2009PercolationNetworks}. 

Undoubtedly, patterns of social contacts drive transmission and they have been studied in the context of respiratory disease in Hong Kong \cite{Leung2017SocialKong}. Elsewhere students have been invited to provide detailed records of contacts ranked by intimacy levels \cite{Edmunds2006MixingDiseases}. Animals have been studied, and connections drawn to their likelihood of contracting parasites or infection diseases \cite{Altizer2003SocialStudies} based on social organization. 

We offer no deeper understanding of human interaction patterns, but we exhibit additional motivation for the functional form suggested by Chowell et al in the early stages of an epidemic. However unlike that adjustment, which is ad hoc, our suggestion takes the form of a delay differential equation compartmental model strongly motivated by a physical agent model. Numerical solution suggests this coincides roughly with Chowell initially but after peak infection, also provides some nuanced dynamics arising from a change in the typical time since infection. Our suggestion is to augment one of the three SIR equations as:
\begin{equation}
\label{eqn:infected_intro}
    \frac{\partial}{\partial t}i(t) = \beta_0 \bar{P}(t) i(t) s(t) - \gamma i(t)
\end{equation}
where $\bar{P}$ is the infected population average of an attenuation function
$$
   Ein(\alpha_0 t) =  \left(\frac{1-e^{-\alpha_0 t}}{\alpha_0 t}\right)
$$
for a parameter $\alpha_0$. Thus $\bar{P}$ might might be estimates as
\begin{equation}
\label{eqn:pbar_intro}
      \bar{P}(t;\alpha) =  \frac{ \int_{s=0}^t Ein(\alpha(t-s)) \overbrace{\left[ \frac{\partial i(s)}{\partial s} +\gamma i(s) \right]}^{new\ infections} e^{-\gamma (t-s)} ds}{\int_{s=0}^t \left[ \frac{\partial i(s)}{\partial s} +\gamma i(s) \right] e^{-\gamma (t-s)} ds } 
\end{equation}
when solving numerically the differential equations. As with any adjustment to an otherwise simple compartmental model this retains a fair degree of convenience, yet it captures important effects that would otherwise call for network, meta-population or agent based models of some variety. These are also valid approaches but not always as easy to wield. 

For short times the average $\beta_0 \bar{P}$ might be quite close to $\beta Ein(\alpha_0 t)$ and therefore we also consider the model where infection takes on this simpler form. In Section \ref{sec:adhoc} we treat the physical model as a suggestion for a very early stage infection attenuation, and examine some analytical properties of the growth. In Section \ref{sec:delayed} we provide a system with better fidelity to the physical model. Both models fall within a broad class of models considered by Kermack and McKendrick \cite{1927AEpidemics}, \cite{Kermack1991ContributionsEndemicity} \cite{Kermack1991ContributionsEndemicityb}. 

However unlike Chowell et al's suggestion, the augmented model we suggest can be viewed as a solution to the physical model which is presented in Section \ref{sec:physics}. Here agents populate the plane and make gaussian jumps to nearby points, thereby colliding with others. The conditional probability that an encounter is novel since time $0$ is given by $Ein(\alpha_0 t)$ and, although we choose to augment the SIR model for simplicity, it should be clear that the same conditional probability could be used to attenuate infection any compartmental model.

There are some advantages to the correspondence between a physical model and an otherwise ad-hoc infection reduction function (even if empirical evidence already provides a strong enough motivation for inclusion of the latter). 
\begin{enumerate}
    \item There is a physical interpretation of $\alpha_0$, as inverse density. 
    \item The physical model is more obviously extensible.
    \item Extensions to the physical model may benefit from knowledge that the special case is solved (as with the use of control variates, for example). 
\end{enumerate}

In regard to the first point, we hasten to add that the model we present in Section \ref{sec:physics} is a highly stylized model. Although there is an interpretation offered in relation to density, there are also obvious ways in which the particles don't really behave like people. They make no effort to avoid each other as density increases, or conversely to seek out contact as we would expect from people living in the countryside. Nonetheless some readers may wish to interpret $\alpha_0$ as related to density in some weaker fashion, and thus the relationship may prove useful.

\section{An ad hoc attenuation of infection rate}
\label{sec:adhoc}

We begin with the simplest augmentation. To model very early stage growth we suggest replacing a constant infection coefficient $\beta_0$ with a declining function of time governed by an additional 
parameter $\alpha_0$. The function finds interpretation as a probability of novel interaction. This adjustment cannot be viewed as equivalent to a physical model - at least not for very long - but it will be further motivated in Section \ref{sec:physics}. The following adjustment will be shown to yield qualitative changes in model behaviour including a transition from exponential to polynomial growth, and ability to fit any power law of growth. We write
\begin{equation}
\label{eqn:beta}
    \beta(t;\beta_0, \alpha_0) =  \beta_0 \overbrace{\left(\frac{1-e^{-\alpha_0 t}}{\alpha_0 t}\right)}^{Ein(\alpha_0 t)}
\end{equation}
where $Ein$ is a naming convention carried over from complex analysis. We will refer to $t \mapsto Ein(\alpha_0 t)$ as the attenuation function and it will
be natural to consider the parameter $\alpha_0$ as the inverse of $\tau_{\alpha_0}$, briefly $\tau$, that represents an important time scale. This time, possibly on the scale of weeks, measures how long it takes for our circle of acquaintances to grow stale. 

Any compartmental model be augmented in this fashion and we choose to illustrate by adjusting the SIR model, described by the following differential equations
\begin{eqnarray}
\label{eqn:sir}
    \frac{\partial}{\partial t} s(t) & = & -\beta_0 \frac{1-e^{-\alpha_0 t}}{\alpha_0 t} s(t) i(t) \nonumber \\
 \frac{\partial}{\partial t} i(t) & = & \beta_0 \frac{1-e^{-\alpha_0 t}}{\alpha_0 t} s(t) i(t) - \gamma i(t) \nonumber \\
  \frac{\partial}{\partial t} r(t) & = & \gamma i(t) 
\end{eqnarray}
where the attenuation function has been inserted. We will see $\beta(t)$ arises endogenously in a continuous spatial model involving an infinite number of agents, which is to say that its aggregate population dynamics are described precisely by equations \ref{eqn:sir}. 

\subsection{Comparison to prior work}

As noted Equation \ref{eqn:sir} form is similar but not equivalent to at least one other proposal for attenuating transmission. Chowell et al define a family of augmented SIR models with a three parameter family where $\beta_0$ plays an identical role to our use. Chowell et al's model differs because two parameters rather than one control its temporal behaviour. 
$$
    \beta(t) =  \beta_0 \left\{ (1-\phi)e^{-qt}+\phi \right\}
$$
In the space of all possible attenuation functions the curve traced out we vary $\alpha_0$ in our model is not nested by the manifold defined by $\phi$ and $q$. However for small values of $t$ a series expansion reveals that our model translates roughly to the choices:
\begin{eqnarray*}
    \phi & =& 1/4  \\
     q  & = & \frac{2}{3} \alpha_0
\end{eqnarray*}
at least for small $t$. In Section \ref{sec:delayed} we will further motivate the form adopted by these authors by showing that infection rate drops quickly and then plateaus, thus justifying the form in Equation \ref{eqn:chowell} with a choice of non-zero $\phi$ (that might typically taking on larger values than $\phi=1/4$).

\subsection{Early stage integrated attenuation and growth rate}

The use of the attenuation function produces a qualitative change in early stage dynamics of $i(t)$, modifying it from exponential to approximately polynomial growth. However this takes a short time to come into effect. Until we start running into people we have already run into before, our model first grows exponentially. 

We distinguish ``initial phase'' from ``early stage'' for this reason, as will be seen, though both may correspond to a period where the susceptible population is largely unchanged. We set $s(t)=1$ in keeping with the usual interpretation of ``early stage growth''. Following the usual development of the SIR model with unit population we have
$$
    \frac{\partial}{\partial t} \log( i(t) ) = \beta(t) - \gamma 
$$
solved by 
\begin{equation}
    \label{eqn:growth}
    i(t) = i(0)e^{-\gamma t} \exp\left( \int_0^t \beta(s)  ds \right)
\end{equation}
and out attention turns to the rate of net growth in the infected population. This will drives the total number of cases and the trajectory of $i(t)$ is very much the story of the epidemic we would like to understand, even if it is a latent variable disguised by testing delays, asymptomatic sufferers and other issues.

With $\beta(t)=\beta_0$ a constant, as with the traditional compartmental models, the growth equation \ref{eqn:growth}
dictates that $i(t)$ will grow exponentially with 
exponent $\beta_0-\gamma$. However with $\beta(t)=\beta_0 Ein(\alpha_0 t)$ properties of the exponential integral function will determine the magnitude of this changing exponent. We have
$$
 i(t) = i(0)e^{-\gamma t} \exp\left( \int_0^t \beta_0 \frac{1-e^{-\alpha_0 s}}{\alpha_0 s}  ds \right) 
$$
and from the vantage point of exponential growth we take interest in the exponent applicable between any two times. After a coordinates $u=\alpha_0 s$ we have 
$$
        g(t_1,t_2) = \frac{\log(i(t_2)/i(t_1))}{t_2-t_1} = \beta_0 \overbrace{\frac{1}{\alpha_0 t_2- \alpha_0 t_1} \int_{\alpha_0 t_1}^{\alpha_0 t_2} \frac{1-e^{-u}}{u}  du}^{growth\ attenuation}    - \gamma
$$
It may already be clear to the reader that a ``bending of the curve'' takes place absent any intervention even if we maintain the assumption $s(t)\approx 1$. The extent to which this integrated attenuation deviates from unity determines how far growth deviates from the SIR benchmark. 

\subsection{Approximations to initial growth}

We will use the notation
\begin{equation}
   E_1(x) = \int_1^{\infty} \frac{e^{-vx}}{v} dv = \int_0^1 \frac{e^{-x/u}}{u} du     
\end{equation}
for the exponential integral function. We have
\begin{equation}
\label{eqn:integral}
   \int_0^{t'} \frac{1-e^{-u}}{u} du = \tilde{\gamma} + E_1(t') + \ln(t') 
\end{equation}
adopting the convention that primed times are intended to play the role of $t{\alpha_0}$, and denoting the Euler-Mascheroni constant by $\tilde{\gamma} \approx 0.57$ to distinguish it from recovery rate $\gamma$.\footnote{In between the similarity of $\tilde{\gamma}$ to $\gamma$ and $Ein$ to $E_1$ (sometimes written $Ei_1$ which confusingly would
bring us even closer to the attenuation function $Ein$) the reader may question our notation choices. However these conventions are standard.} For $t'<1$, which corresponds to times $t<\tau_{\alpha_0}$ we can use the convergent sequence
$$
\tilde{\gamma} + E_1(t') + \ln(t') =
t'-{\frac{1}{4}}{t'}^{2}+{\frac{1}{18}}{t'}^{3}-{\frac{1}{96}}{t'}^{4}+{
\frac{1}{600}}{t'}^{5}+O \left( {t'}^{6} \right) 
$$
to establish that for $t \ll \tau$ there is no appreciable growth attenuation. We start out with instantaneous growth given by 
$\beta_0-\gamma$ as with the SIR model. However this expansion also makes clear that if we look at mean growth $g(0,\tau)$ over the first period up to time $\tau=1/\alpha_0$ then, setting $t'=1$, we have a reduction of growth of approximately $20$ percent. 

Although the power series is convergent only for $t'<1$ the formula \ref{eqn:integral} is valid for all time. Table \ref{tab:numbers} lists growth rate approximations relative to the attenuated SIR model, assuming $s(t)\approx 1$. We use standardized times in units of  $\tau=1/\alpha_0$

\begin{table}
 \label{tab:numbers}
\begin{center}
    \begin{tabular}{|l|l|l|l|l|l|l|l|l|l|l|l|}
\hline 
      $t_1'$ & $t_2'$ & $g(t_1',t_2')$  & $t_1'$ & $t_2'$ & $g(t_1',t_2')$ & $t_1'$ & $t_2'$ & $g(t_1',t_2')$ & $t_1'$ & $t_2'$ & $g(t_1',t_2')$\\
     \hline 
        0  & 1   & 0.795   & 1 & 2 & 0.522   & 0 & 10 & 0.288   & 100 & 101 & 0.010 \\
        0 & 2    & 0.659   & 2 & 3 & 0.370   & 10 & 20 & 0.069  & 200 & 201 & 0.005 \\
        0 & 3.   & 0.562  & 1 & 3. & 0.446   & 10 & 11 & 0.0953 & 1000 & 1001 & 0.001 \\ 
        \hline 
\end{tabular}
\end{center}
\caption{Table of early phase growth rates relative to SIR model as predicted by Equation \ref{eqn:integral}. This assumes, however, that the susceptible population has not been materially
diminished - something that is unlikely to be the case for the larger values shown.}
\end{table}

\subsection{Polynomial growth}

Morally speaking, only the $1/u$ term in the integral will be important in the middlegame when $t > \tau$, assuming our key assumption takes us that far. If it is the case that herd immunity has not set in and we still have $s(t)\approx 1$, then the integral will increase roughly as the logarithm of time, behaving like 
$$
    i(t_2) \approx i(t_1) e^{-\gamma (t_2-t_1)} e^{\frac{\beta_0}{\alpha_0} (\log(t_2)-\log(t_1))} = e^{-\gamma (t_2-t_1)} \left(\frac{t_2}{t_1}\right)^{\frac{\beta_0}{\alpha_0}}
$$
Growth resembles not an exponential but a power law damped by an exponential. The exponent is 
$$
     p = \frac{\beta_0}{\alpha_0}
$$
which suggests a way to calibrate both $\beta$ and $\alpha$ using two equations instead of one.\footnote{A slightly different approach might benefit from historical experience (panel data) if that exists, because the declining stage of an epidemic might give us a handle on $\gamma$ directly. This can then be used to adjust the data, multiplying it by $e^{\gamma t}$ to obtain a quantity that should be proportional to a power law of time. The pair $\beta$ and $p$, rather than $\beta$ and $\alpha$, may even be a more natural choice of parameters.}

\subsection{Peak infection may be unrelated to herd immunity}

A situation where these relatively short time approximations may be beneficial occurs when an infected population arrives in a new city. For example passengers may disembark from a cruise ship or arrive on a long haul flight. Their novelty clocks are reset to $t=0$, as it were. Initially the entire cohort of infected people undergoes rapid decline in novelty probability and it is possible that this drag on growth can contribute to a rapid extinction of the virus, even when we have 
$$
  R_0 := \frac{\beta_0}{\alpha_0} > 1 
$$
which would, in the original SIR model, lead to exponential growth all the way until herd immunity. Instead we can observe that once we pass a given time (say $t=7$ for concreteness) the term $e^{-u}$ is bounded above and below for the duration of the integral, so we only have $1/t$ do deal with. For example we might fix an integer $k$ an note that
$$
       (1-e^{-k\tau}) \int_{u=k\tau}^{(k+1)\tau } \frac{1}{u} du    \le    \int_{u=k\tau}^{(k+1)\tau } \frac{1-e^{-u}}{u} du \le (1-e^{-(k+1)\tau}) \int_{u=k\tau}^{(k+1)\tau } \frac{1}{u} du
$$
showing that the attenuation between times $k\tau$ and $(k+1)\tau$ is bounded above an below. Translating back into
the population of infected these bounds look like
\begin{eqnarray*}
  i((k+1)\tau) & > & i(k\tau) e^{-\gamma \tau} e^{\frac{\beta_0}{\alpha_0} (1-e^{-k\tau}) \int_{u=k\tau}^{(k+1)\tau } \frac{1}{u} du  }  \\
               & > & i(k\tau) e^{-\gamma \tau} e^{\frac{\beta_0}{\alpha_0} (1-e^{-k\tau}) \log(\frac{k+1}{k} ) } 
\end{eqnarray*}
with the net growth rate during this period exceeding
$$
         \frac{\beta_0}{\alpha_0} \frac{1-e^{k\tau}}{\tau} \log(1 + 1/k ) -\gamma
$$
and less than 
$$
        \tau \frac{\beta_0}{\alpha_0} \frac{1-e^{(k+1)\tau}}{\tau} \log(1 + 1/(k+1) ) -\gamma
$$
The bounds squeeze the growth rate towards 
$$
 \frac{\beta_0}{\alpha_0} \frac{\log(k)}{k}-\gamma 
$$
as $k \rightarrow \infty$, which is to say that as a function of time, the growth rate approaches
\begin{equation}
\label{eqn:exponent}
   \frac{\beta_0}{\alpha_0} \frac{\log(t/\tau)}{t/\tau}-\gamma 
\end{equation}
where again $\tau=1/\alpha$ is the forgetting time. Clearly the expression \ref{eqn:exponent} does not remain positive. Even before we account for the fact that $s(t)=1$ might not be true anymore, which as been our working assumption in these calculations, it is clear that a critical time will arrive when this quantity is zero. There is an intuitive way to represent that moment if we write $p=\frac{\beta_0}{\alpha_0}$. Rearranging we can express zero net growth in dimensionless time $t'$ as  
$$
        e^{\gamma t'} = (t')^p 
$$
We rearrange to find 
$$
        t' = e^{\frac{\gamma}{p}t'}
$$
whose solution, assuming there is one, is given by a branch of the Lambert W function  
$$
    t' = - \frac{p W_n\left(-\frac{\gamma}{p} \right)}{\gamma}
$$
Interestingly the point of peak infection is also recognized as the solution to an equation of the form
$$
     \log(x) = c x
$$
studied by Euler. In this way the battle between polynomial growth in new infections and exponential decline due to recovery turns may come to an early turning point. In Section \ref{sec:delayed} we shall see that in a more strongly motivated model this battle must be fought and won quickly, however, because the rapid decline in novelty will not last. 

\subsection{Extinction}

In the event that the virus is nipped in the bud with help from a kind of local herd effect, as we might characterize it, we enter decline. It may well be the case that the decline is somewhat different to the standard model because new infection growth has stalled (contribution from new infections will fall quickly towards a low multiple of the recovery coefficient $\gamma$). It follows that decline will be a mostly straight line in log space with a highly predictable slope largely independent of conditions that drive contagiousness.

Also in the event that attenuation in infection rate continues to be the most important dynamic (i.e. $s(t)\approx 1$ and herd immunity is still far off) we can compute an approximate count of the total recovered. This is an Euler integral of the second kind
$$
  r(\infty) = \gamma \int_1^{\infty} i(s) ds \approx i(1) {\gamma}^{-{\frac {\beta}{\alpha}}}\Gamma \left( {\frac {\beta+
\alpha}{\alpha}} \right) 
$$
where $\Gamma$ is the Gamma function. It is also possible to compute the integral starting from $0$ instead of $1$ but the result is a much more complicated combination of special functions. Since $p=\frac{\beta}{\alpha}$ is the approximate power law
we note that for integer $p$ the ratio of total recovered to infected at time $t=1$ is approximately factorial in the power law and a power law of the mean recovery time $\tau=1/\gamma$.  
$$
   \frac{r(\infty)}{i(1)} \approx \frac{p!}{\gamma^p} = p! \tau^p 
$$
For example if growth is quadratic then, compared with the total reached at time $1$, the total number of infected eventually grows by a factor of $8$. However, we would emphasize that even the middlegame can be quite different, as we shall discuss next, so the domain on which these endgame formulas are valid is likely to be limited. 

\section{Delayed differential equation model}
\label{sec:delayed}

The suggestion in Section \ref{sec:adhoc} is made in the interest of simplicity.  We have seen that the first order effects can be understood analytically to some extent, assuming $s(t) \approx 1$. That said, the form is not motivated in a satisfying manner. In this section we shall consider a close cousin that, while not representing itself as convenient in quite the same sense, bears much closer fidelity to the physical model in Section \ref{sec:physics} and can, it stands to reason, perhaps be leaned on more heavily for middle and endgame analysis. 

This model falls into a class of compartmental models where infection rate is a function of ``vintage'', which is to say how long someone has been infected for. 

\subsection{Cohort mean novel collision probability}

Once again we replace the constant infection rate $\beta_0$ in the SIR model and leave everything else unchanged. We shall write
\begin{equation}
\label{eqn:infected}
    i'(t) = \beta_0 \bar{P}(t) i(t) s(t) - \gamma i(t)
\end{equation}
for the infection equation where $\bar{P}(t;\omega)$ depends on history $\omega$ of $i(t)$. It is intended to be an approximation to the mean probability of novel encounter averaged over every person who is infected. The other two equations are as one would expect. The susceptible population is depleted $
  s'(t) = -\beta_0 \bar{P}(t;\alpha) i(t) s(t)
$
and the recovered population grows
$
r'(t) = \gamma i(t)
$
as before. The function $\bar{P}$ is a weighted average of $Ein$
\begin{equation}
\label{eqn:pbar}
      \bar{P}(t;\alpha) =  \frac{ \int_{s=0}^t Ein(\alpha(t-s)) \overbrace{\left[ \frac{\partial i(s)}{\partial s} +\gamma i(s) \right]}^{new\ infections} e^{-\gamma (t-s)} ds}{\int_{s=0}^t \left[ \frac{\partial i(s)}{\partial s} +\gamma i(s) \right] e^{-\gamma (t-s)} ds } 
\end{equation}
where for convenience we repeat the definition 
$$
   Ein(u) = \frac{1-e^{-u}}{u}
$$
The calculation of $\bar{P}$ may be viewed as an application of Bayes Rule.\footnote{For a discussion of the general class see Kermack and McKendrick See pp 704 at https://royalsocietypublishing.org/doi/pdf/10.1098/rspa.1927.0118} The probability that a person who was infected at time $s$ still remains in the set of infected people at time $t$ provides the term $e^{-\gamma(t-s)}$, and we also weight by the number of people who entered the infected cohort at that time. Alternative expressions may be obtained integrating by parts. A solution by means of interactive approximations is considered by Kermack and McKendrick \cite{Kermack1991ContributionsEndemicityb}. Our examination is driven by numerical solution of the delay differential equations. 

\subsection{Vintage and momentum}

The behaviour of $\bar{P}(t)$ near $t=0$ may be reasonably be approximated by $Ein(t)$ since the vintage of the infected group will be dominated, at least momentarily, by the initially infected people. The previous discussion in Section \ref{sec:adhoc} applies (as do analytic results in Kermack and McKendrick \cite{Kermack1991ContributionsEndemicityb}). 

As one might anticipate, the mean probability of novel encounter $\bar{P}$ will initially fall. 
However once the vintage of the infected group has grown somewhat, $\bar{P}$ may tend to plateau or decline much more slowly until such time as the course of the epidemic starts to alter the makeup of the group more dramatically. In the declining phase relatively few new infected people enter the cohort and the average time spent by anyone who is infected rises. This makes the attenuation effect of lack of novelty even more important, and may predict a downward path of infections that is more abrupt than expected in a classical compartmental model. 

The movement of the typical vintage of an infected person back and forth is a form of momentum. In the early phase new people entering the pool of infected have very little attenuation since everyone they meet is novel (as judged from the time they are infected), thus leading to a value of $\bar{P}$ closer to $1$ than in a steady state. Correspondingly, when this dynamic reverses $\bar{P}$ will fall. This effect feeds back into the creation of fewer infections, thus further back-weighting of the typical vintage. 
\begin{figure}
    \centering
    \includegraphics[scale=0.6]{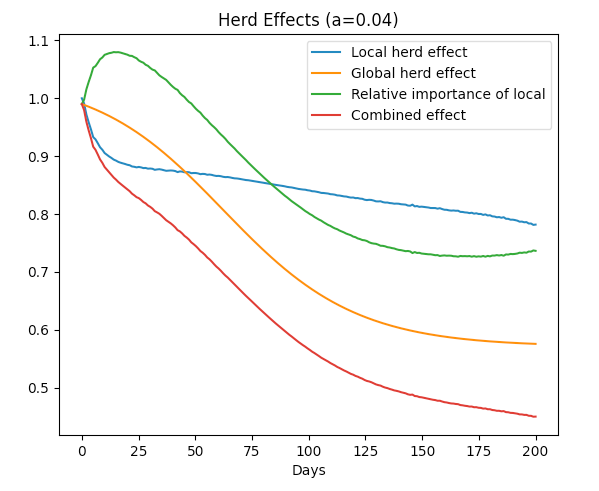}
    \caption{The mean attenuation $\bar{P}$ is shown and interpreted as a local herd effect. Also shown is the percentage of susceptible people (the global herd effect). In this example $a=0.04$ corresponds to a $37\%$ diminution in infection over $25$ days due to staleness of the circle of acquaintance. This effect is initially more important than the drawing down of the susceptible population. Only after $50$ days does the impact of the herd effect (in the usual sense) contribute equally.}
    \label{fig:004}
\end{figure}

\begin{figure}[h!]
    \centering
    \includegraphics[scale=0.4]{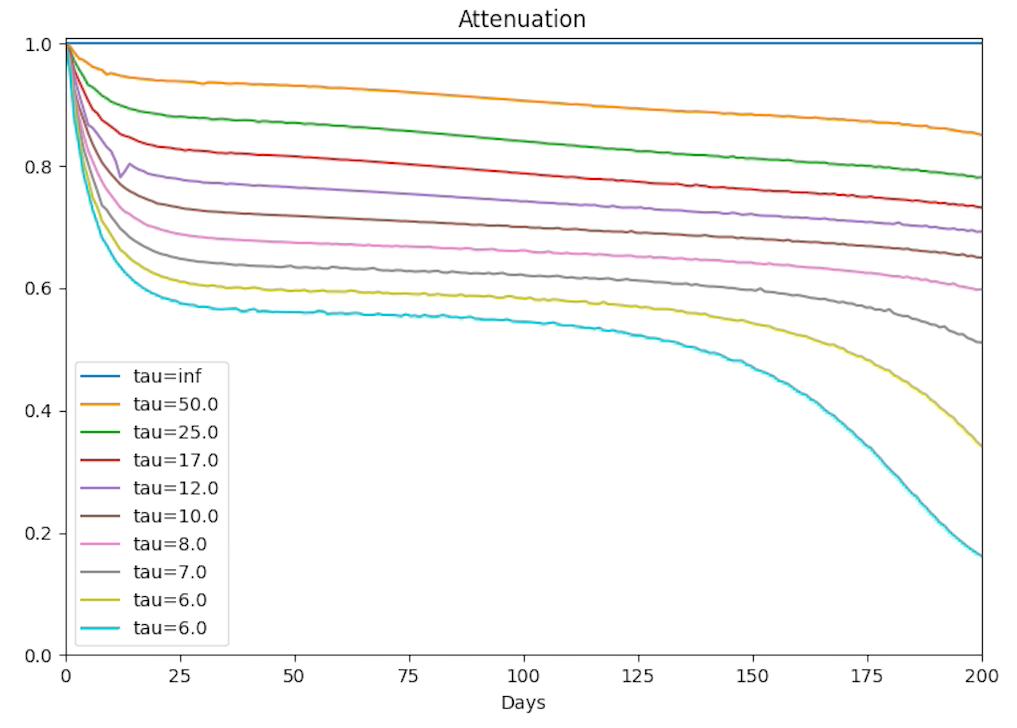}
    \caption{Trajectories of $\bar{P}$ for difference values of $\alpha_0=1/\tau$ showing the pronounced momentum effect. Slowing of the number of new infections relative
    to recovery shifts the average time of infection back relative to the present, making the cohort of infected people less dangerous to the susceptible population.}
    \label{fig:attenuation}
\end{figure}

Figure \ref{fig:004} provides an example of the path of $\bar{P}$ during the rise and fall of the infected population, and its relative importance compared with the herd effect of falling $s(t)$ that would, in the absence of the attenuation due to novelty, be the only break on the epidemic's rise. Figure \ref{fig:attenuation} shows trajectories of $\bar{P}(t)$ for varying values of $\alpha_0$. All display a characteristic initial phase followed by a gentle decline as the vintage shifts. In the scenario where $\tau$ is less than a week (and also less than $1/\gamma=10$) this dynamic is quite pronounced, leading to a situation where almost $80$ percent of interactions are not novel.\footnote{Declining novelty of interaction may be viewed as partial recovery. The attenuation function is not exponential but closer to the inverse of time expired, so novelty is not exactly equivalent to an increase in the recovery rate. However the exponential function is probably a mediocre approximation to recovery to begin with, so some tightening of this connection should be possible if we allow ourselves some further flexibility in the SIR prescription.} 

\begin{figure}[h!]
    \centering
    \includegraphics[scale=0.4]{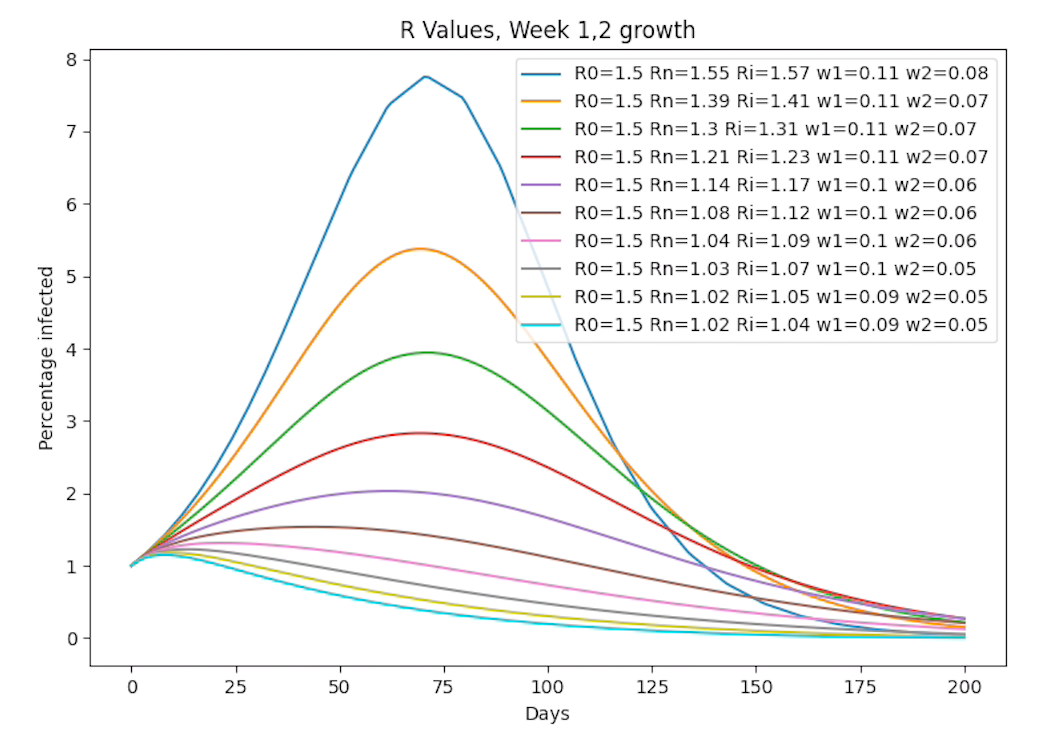}
    \caption{Infection trajectory $i(t)$ for different values of $\alpha_0$ but identical $\beta_0$, $\gamma$ and hence $R_0=\beta_0/\gamma$. The quantity 
  $R_n=1/s(t^*)$ where $t^*$ is the time of peak infection and also  $R_i=-\frac{log(\frac{1-r}{s(0)}}{AR-(1-s(0)}$ are shown. Both these ex-post estimates
  of reproduction roughly coincide with $R_0$ for the base case $\alpha_0=0$ but differ markedly for other values. \cite{Obadia2012TheOutbreaks}. Attenuation of infection due to lack of novel interactions has little effect in the first week ($w_1$) but is noticable in the second ($w_2$). The difference between the top and bottom trajectories amounts to roughly a ten fold difference in total case count and peak infection.}
    \label{fig:rvalues}
\end{figure}

\begin{figure}
    \centering
    \includegraphics[scale=0.4]{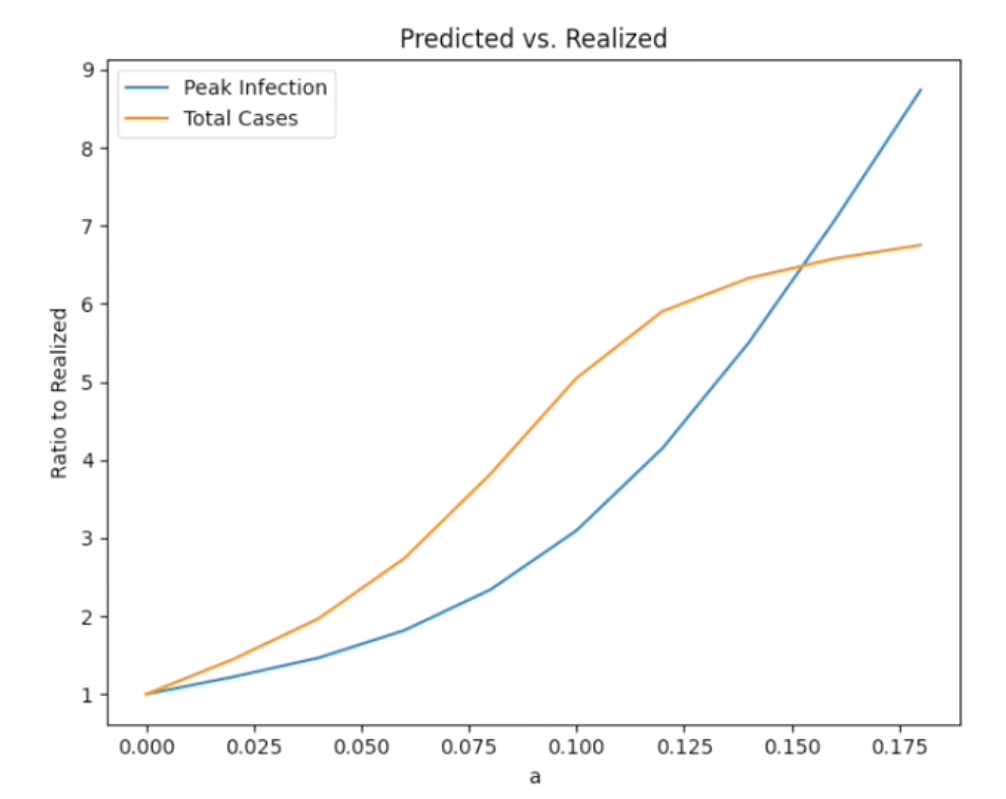}
    \caption{The extent to which a na\"ive estimate of $R_0$ from the first week of data would overestimate final case count, shown for various values of $\alpha=1/\tau$. }
    \label{fig:overestimate}
\end{figure}

\subsection{Visualizing two possible turning points}

\begin{figure}
    \centering
    \includegraphics[scale=0.65]{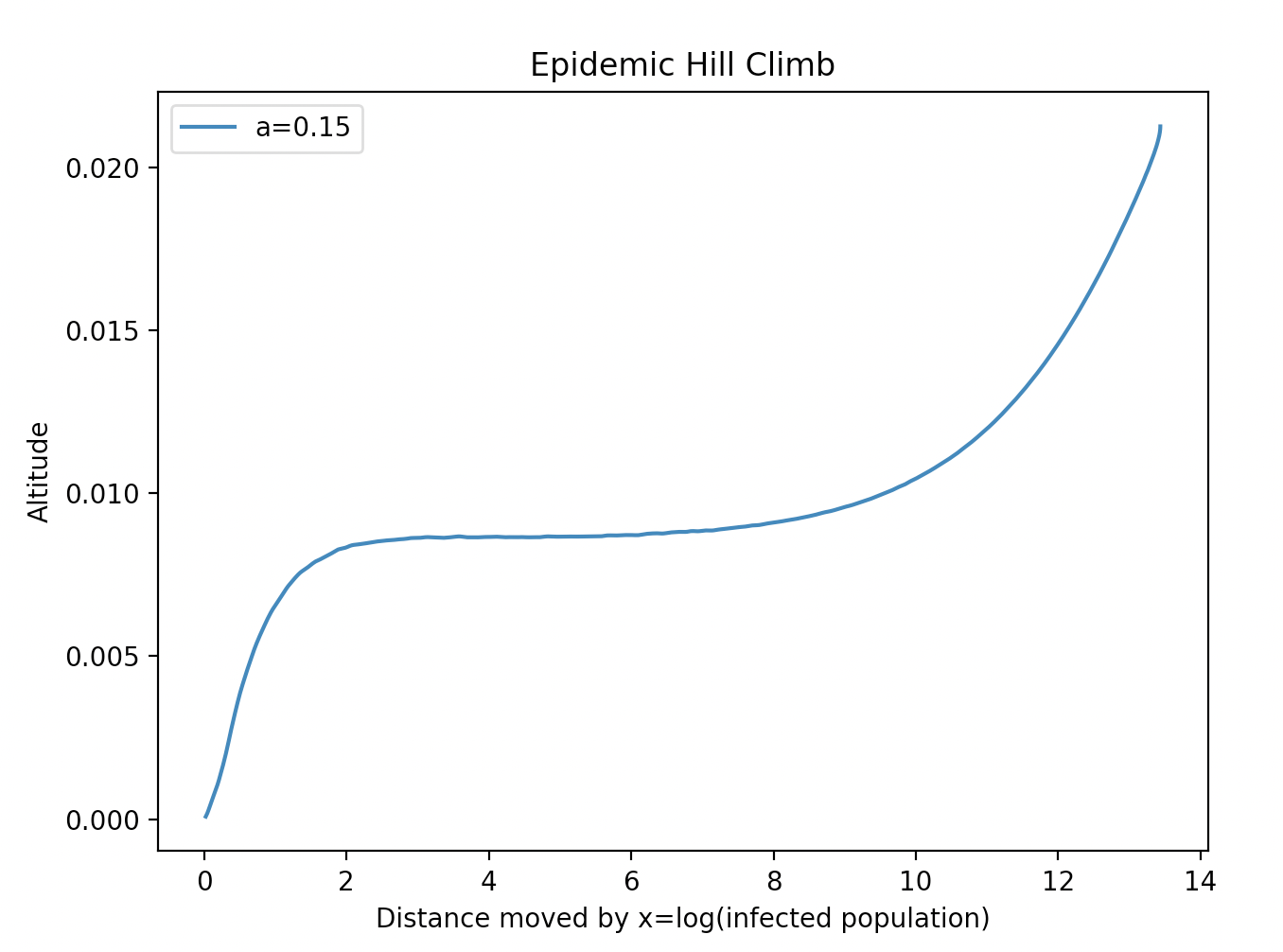}
    \caption{Hill climbing visualization of the forces acting on the logarithm of infected population. Based on numerical simulation of the delay differential equations with repeat contact infection attenuation, we compute numerically the acceleration of $x=\log(i(t))$. Then using elementary Newtonian mechanics we reverse engineer a hill that would create the exact same acceleration for a sliding object - with initial velocity calculated so that the object comes to rest at the onset of herd immunity. In stark contrast to the classic constant parameter SIR model there are two hills to climb rather than one. It is possible that an epidemic, initiated by the arrival of an infected population by boat or plane for example, dies out very quickly. However once that first hill is crested there is little to slow growth until herd immunity sets in. This picture strongly suggests aggressive early intervention to take advantage of the temporary phase where infection attenuation due to shrinking circles of novel acquaintance is greatest, because after this stage we receive relatively little assistance from the effect until much later when peak infection has come and gone. This picture also suggest that bimodal outcomes for countries are more likely than they are in the classic model, and it may help explain the distribution of cases as we look across countries like Australia and New Zealand who were able to quickly control growth, versus others such as Italy, the United States of Sweden who were not.}
    \label{fig:hill}
\end{figure}

To illustrate the qualitative distinction between the model suggested and the constant parameter SIR model, we introduce what may be a novel way of visualizing the course taken by the epidemic. Figure \ref{fig:hill} 

\section{A continuous spatial model for interactions}
\label{sec:physics}

We now exhibit a physical model for disease spread motivating the model described in Section \ref{sec:delayed} and to a lesser extent the model provided in Section \ref{sec:adhoc}. We wish to capture the impact of repeated contact without using a multi-population model, network models or finite agent based modeling approach. We suggest instead a modeling framework in which a uniform homogeneous population populates every point on the plane. There are different ways to present this but here we will adopt a style reminiscent of first principles calculus so that at any point in our argument the model can be considered an agent model, albeit one with an arbitrarily large number of agents whose number is proportional to $1/\delta^2$.  

Home locations of individuals are modeled as standard normally distributed random variables with position $\mu=(\mu_x,\mu_y) \in \mathcal{R}^2$. Fix $\epsilon \ll 1$, $\delta \ll 1$. We assume that on $n_{\epsilon,\delta}(\beta_{0})=\frac{\beta_{0} \delta^2}{4 \epsilon^2}$ occasions per unit time the individual's position is drawn from his or her distribution, namely
\begin{equation}
\label{eqn:rho}
      \rho_{\mu}(x,y) = \frac{1}{2\pi} e^{- (x-\mu_x)^2/2- (y-\mu_y)^2/2}
\end{equation}
Individuals are spaced a distance $\delta$ apart on a square lattice tiling the plane. In what follows the reader may wish to consider $\epsilon \ll \delta \ll 1$ so that the distinction between $n_{\epsilon,\delta}(\beta_{0})$ and its nearest integer can be swept under the rug. 

\subsection{Relationship to Ornstein-Uhlenbeck walk}

The model is in some respects similar to the Ornstein-Uhlenbeck motion model proposed in \cite{oupandemic}. That model suggests people follow regime switching Ornstein-Uhlenbeck walks as they proceed from home to work, and suggests varieties of home and work location generation. 

Here we specialize to what might be called lockdown conditions (no work locations or commuting) and we turn on strobe lighting, as it were. To tighten the connection:
\begin{enumerate}
    \item Homes are uniformly spaced, with density $1/\delta^2$. 
    \item The ergodic mean squared distance from home is unity. This amounts to setting $\sigma^2=2\kappa$ in the Ornstein-Uhlenbeck walk.
    \item Both $\sigma$, $\kappa$ are taken towards $\infty$ which retaining a constant ratio $\sigma^2/\kappa$. 
\end{enumerate}
Recall that a particle at position $x$ following a univariate process $dx_t = -\kappa x_t + \sigma dW_{x,t}$ pulled to the origin will, after elapsed time $t$, arrive at normally distributed position with mean and variance given by 
\begin{eqnarray}
\label{eqn:law}
     \mu_x(t) & = &  x e^{-\kappa t} \rightarrow 0 \nonumber \\
     \sigma_x^2(t) & = & \sigma^2 \frac{1-e^{-2\kappa t} }{2\kappa} \rightarrow 1
\end{eqnarray}
This follows from direct solution of the stochastic differential equation, and the analogous results holds for the $y$-coordinate also when $(x_t,y_t)$ is a two dimensional Ornstein-Uhlenbeck walk.\footnote{The combination of independent walks can be considered a random walk subject to a pull proportional to the distance from the origin since the force of strength $\kappa \sqrt{x^2+y^2}$ decomposes into components of respective strength $\kappa x$ and $\kappa y$.} It follows that if $\kappa \rightarrow \infty$ and $\sigma \rightarrow \infty$ while simultaneously $\sigma^2/\kappa \rightarrow 2$ then the location of the particle
will converge on a standard bivariate normal distribution, thus justifying equation \ref{eqn:rho} as an approximate model for collisions. 

It is worth noting that other mathematical tools might be brought to bear on the problem of computing collision probabilities for disks following continuous paths in the plane. However a finite sampling of positions is the more relevant analysis if our goal is relating compartmental models to simulation models {\em likely to be employed}. For example: 
\begin{enumerate}
    \item Use equations \ref{eqn:law} to update particle positions
    \item Positions are geohashed to one dimension (via interleaving of binary representations as with a space filling curve)
    \item Membership of a set of cells occupied by infected people is checked in $O(1)$ time.
\end{enumerate}
The calculus we present is intended to bridge the gap between this kind of efficient agent based motion model and compartmental models - modulo the use of the limit $\kappa \rightarrow \infty$. A second order analysis relevant to smaller values of $\kappa$ may be possible using the fact that equation \ref{eqn:law} betrays the degree to which sequential points deviate from independent draws. 

\subsection{Collision probabilities}

Without further motivation we now assume motion occurs as independent draws from a standard normal distribution. We consider two people with different homes, which is to say different mean
locations. Let $\mu'$ denote the location parameter for a second distribution:
$$
      \rho_{\mu'}(x,y) = \frac{1}{2\pi} e^{- (x-\mu'_x)^2/2- (y-\mu'_y)^2/2}
$$
Collision probability near $(x,y)$ is proportional to the product
\begin{eqnarray}
\rho_{\mu}(x,y)\rho_{\mu'}(x,y) & = & \frac{1}{4\pi^2} e^{- (x-\mu'_x)^2/2- (y-\mu'_y)^2/2 - (x-\mu_x)^2/2- (y-\mu_y)^2/2} \\
                & = & \frac{1}{4\pi^2} e^{- (x- \frac{\mu_x+\mu'_x}{2})^2/2- (y-\frac{\mu_y+\mu'_y}{2})^2/2 - (\frac{\mu_x-\mu_x'}{2})^2/2 - (\frac{\mu_y-\mu'_y}{2})^2/2} 
\end{eqnarray}
Moreover if we integrate across all possible collision locations, the probability that the first person falls within a square area $\epsilon^2$ centered 
on the second person's location, denoted $P(<\epsilon)$ approaches
\begin{eqnarray}
  \frac{  P(<\epsilon)}{ \epsilon^2}  & \rightarrow &\iint_{\mathcal{R}}
        \rho_{\mu}(x,y)\rho_{\mu'}(x,y) dx dy \\
    & = & \frac{1}{2\pi} e^{-d^2/8}
\end{eqnarray}
where $d^2=(\mu_x-\mu_x')^2 +(\mu_y-\mu_y')^2$ is the squared distance between them. Any two people whose homes are 
a distance $d$ apart will collide on
$$
       n_{\epsilon}(\beta_{0})P(<\epsilon) = \frac{\beta_{0} \delta^2}{2\epsilon^2} \cdot  \frac{\epsilon^2}{2\pi} e^{-d^2/8} = \frac{\beta_{0} \delta^2}{4}\frac{1}{2 \pi} e^{-d^2/8}
$$
occasions per unit time. 

\subsection{Probability of any collision}

Intuitively, because a person's random locations are likely to fall closer to the origin than far away, they are more likely to have more frequent collisions with those whose homes are closer to them. 

Without loss of generality consider a person centered at the origin $\mu=(0,0)$. Let $\Omega_{r,dr}$ denote the set of points whose distance to the origin lies between $r$ and $r+dr$ for some $dr \ll 1$. There are $\frac{2\pi r dr}{\delta^2}$ denizens of $\Omega$, which is to say people whose mean locations are approximately $r$ from the origin. Therefore the mean number of collisions per unit time, which we might denote  $P( \Omega_{r,dr} )$, is
$$
   P( \Omega_{r,dr} ) = \frac{2\pi r dr}{\delta^2} \cdot \frac{\beta_{0} \delta^2}{4}\frac{1}{2 \pi} e^{-r^2/8} = \frac{\beta_{0}}{4} e^{-r^2/8} r dr 
$$
and by integration the mean number of collision per unit time with anyone 
is $\beta_{0}$. This reveals our intent when setting
$$
   n_{\epsilon,\delta}(\beta_0) = \frac{\beta_0 \delta^2}{4 \epsilon^2}
$$
as the number of occasions per unit time that particles jump to a new location. The parameter $\beta_0$ can be viewed as a mean infection rate if it is assumed that every collision leads to transmission. 

\subsection{Probability of novel collision}

Suppose we multiply the probability that a the person who resides at the origin collides with {\em anyone} living in $\Omega_{r,dr}$ by 
the probability that there was no previous collision with {\em that particular person} recently. The events are conditionally independent so 
this yields the probability of the particle experiencing a novel collision with {\em someone} who lives in the anulus. We can express this as a count of novel collisions occuring between $t$ and $t+dt$ that are unique since time $t=0$, as follows:
$$
  Q(\Omega_{r,dr}) dt =   \overbrace{ \frac{\beta_{0}}{4} e^{-r^2/8} r dr dt}^{P( \Omega_{r,dr} ) dt }    \cdot  \underbrace{ exp\left(-t \frac{\beta_{0} \delta^2}{4}  \frac{1}{2\pi} e^{-r^2/8} \right)}_{no\ previous\ collision } 
$$
Thus the rate at which a given person experiences {\em any} novel collision per unit time at {\em any} distance is obtained by integrating over all anuluae. 
\begin{eqnarray*}
\label{eqn:q}
  Q(t,dt) & \rightarrow  &\int_{r=0}^{\infty} exp\left(-t\frac{\beta_{0} \delta^2}{4} e^{-r^2/8} r dr \right) \cdot \frac{\beta_{0}}{4} e^{-d^2/8} r  dr \\
   & = &  8 \pi  \frac{1-e^{- \frac{\beta_{0} t \delta^2}{8 \pi}   }}{\delta^2 t} \rightarrow  \beta_{0} \ as\ t \rightarrow 0
\end{eqnarray*}
This converges to $\beta_{0}$ for small times $t$ as we would expect given the interpretation of $\beta_0$ as the initial rate. However we see that
$\beta(t)$ converges to zero for large $t$ and therefore, depending on the rate $\beta_0 \delta^2$, we might expect quite an early reduction in growth. 

\subsection{Relationship to compartmental models}

The development above (or Equations \ref{eqn:q} in the next section) can serve as an approximation to the Ornstein Uhlenbeck simulation model. The approximation will
be best when $\kappa$ exceeds the inverse of the typical simulation time step. Alternatively, the random sampling model and the distribution \ref{eqn:rho} can be taken as an ansatz and a model in and of itself.  

Either way, let us next suppose that a motion model is extended so as to constitute a dynamic model for contagion, by means of introducing health states assigned to particles together with Poisson transitions from one state to the next. If these choices are made in a manner that preserves homogeneity and in a manner that introduces no coupling between progression and location (such as the introduction of a hospital at the origin or some such symmetry breaking), then it may well be the case that the continuous agent model's population statistics are described by a compartmental model. 

In our stylized spatial model the state of the population living on $\mathcal{R}^2$ is always homogeneous. An individual particle's history of previous collisions is the only state that does not reset in between jumps. Our calculation of the mean probability of novel encounter may be interpreted as a spatial or population average. Thus, making an informal appeal to the Central Limit Theorem, and in this way the physical model relates to a compartmental model. 

\subsection{A receptor model with continuous limit}

Notice that in the infinite agent model as $\delta \rightarrow 0$ we find so many people living inside the typical radius of excursion that the probability of a collision being unique is arbitrarily close to one. The anonymity of a big city is a real thing, but some modification is necessary if we wish to find a limiting case that doesn't collapse back to the starting point of zero attenuation in infection. 

One way out of this scaling conundrum is to imagine that particles do not represent individual people but instead, a tiny part of a person (such as a receptor). It is our receptors that choose the locations and what matters is not a collision with a unique receptor but a collision with a receptor from a new, unique person that we haven't already infected. 

We can scale up the number of receptors, placing their homes a distance $1/\delta$ apart, but we can assign them to individuals so that a given individual possesses
$$m_{\alpha,\delta} = \frac{8 \pi \alpha_0}{\beta_0 \delta^2}$$
receptors. This means that $\alpha_0$ plays the role of inverse density. There will be $\frac{\beta_0}{8 \pi \alpha_0}$ people per unit area. Recalling that the ratio of $\beta$ to $\alpha$ can play the role of exponent in the polynomial growth stage, this makes for a connection between power laws and density. 

Now let us repeat the thought experiment we have already carried out with a receptor playing the role of a person. The chance of no previous collision with any receptor belonging to an individual is the product of survival probabilities. There will be $\frac{2 \pi r}{\delta^2}$ receptors in $\Omega(r,dr)$ so the probability of a given receptor colliding with another receptor is the same as the probability we previously calculated for a person colliding with another (we are now spacing the receptors by $\delta$, not the people). However each person has $\frac{8 \pi \alpha_0}{\beta_0 \delta^2}$ receptors so this multiplies the odds. The infection probability between $t$ and $t+dt$ is now: 
\begin{eqnarray}
  \tilde{Q}(\Omega_{r,dr}) dt & = &  \overbrace{ \frac{8 \pi \alpha_0}{\beta_0 \delta^2} \cdot \frac{\beta_{0}}{4} e^{-d^2/8} r dr dt}^{any\ collision }    \cdot   
       \underbrace{ exp\left(-t \frac{\beta_{0} \delta^2}{4}  \frac{1}{2\pi} e^{-r^2/8} \right)^{    \overbrace{ \frac{8 \pi \alpha_0}{\beta_0 \delta^2}   }^{m_{\alpha,\delta}}      }   }_{no\ previous\ group\ collision } \\
        & = &   \frac{2 \pi \alpha_{0}}{\delta^2} e^{-d^2/8} r dr dt    \cdot   exp\left(-t \alpha_0 e^{-r^2/8} \right)
\end{eqnarray}
which when we integration over all anulae yields the expected number of infections per unit time 
\begin{eqnarray}
\int_{r=0}^{\infty} \tilde{Q}   dr & = & \int_{r=0}^{\infty} \frac{2 \pi \alpha_{0}}{\delta^2} e^{-d^2/8} r  dr  \cdot   exp\left(-t \alpha_0 e^{-r^2/8} \right) dr \\
        & = & \frac{8 \pi}{\delta^2} \cdot \frac{1-e^{-\alpha_0 t}}{t} \\
         & = & \frac{\beta_0}{\alpha_0} m_{\alpha} \cdot  \frac{1-e^{-\alpha_0 t}}{t}
\end{eqnarray}
However note that in this calculation $\beta_0$ plays the role of its baseline infection rate. This we did to allow we have allow a receptor to play the role of a person. But if we wish to define $\beta_0$ instead as relating to the baseline infectivity per person then we must divide by the number of receptors. With this convention we are back in line with the use of $\beta$ as traditionally used in compartmental models and we find: 
$$
 \beta(t) := \frac{1}{m_{\alpha}} \int_{r=0}^{\infty} \tilde{Q} dr = \beta_0 \frac{1-e^{-\alpha_0 t}}{\alpha_0 t}
$$
where as  noted, $1/\alpha_0$ is proportional to density of people (not receptors). In a low density environment the exponential killing of $\beta(t)$ kicks in quickly. In a high density setting it takes longer to have effect. 

\section{Summary}

We have exhibited a physical model for disease spread in which an attenuation of infection rate due to repeated contact takes on a simple form. This provides stronger motivation for empirically motivated adjustments to compartmental models in which infection rate is attenuated over time in order to reproduce sub-exponential growth. However, unlike previous suggestions, we have shown that the functional form of attenuation arises endogenously in an agent model - a connection that may have other uses. 

Important qualitative differences examined include:
\begin{enumerate}
    \item Two different kinds of turning point, rather than one. 
    \item Transition from exponential to polynomial growth, and back. 
    \item Late stage infection attenuation due to a vintage effect. 
\end{enumerate}
It is worth noting that the model may also react differently to interventions. For example a lifting of a lockdown may lead to a more dramatic change to infections than in a constant parameter epidemic model, due to a partial reset of the ``novelty clock''. 

\bibliography{references.bib,unpublished.bib}
\bibliographystyle{plain}

\end{document}